\documentclass[runningheads]{llncs}
\usepackage{graphicx}
\usepackage{amsmath,amssymb} 
\usepackage{color}
\usepackage{pifont}
\usepackage{multirow}
\usepackage[width=122mm,left=12mm,paperwidth=146mm,height=193mm,top=12mm,paperheight=217mm]{geometry}

\renewcommand {\thetable} {\arabic{table}}
\usepackage{caption}
\usepackage{subfigure}
\usepackage{verbatim}

\begin{document}
\def\ECCVSubNumber{}  

\title{Spatial-Angular Interaction for Light Field Image Super-Resolution}

\titlerunning{LF-InterNet}
\authorrunning{LF-InterNet}
\author{Yingqian Wang$^{1}$, Longguang Wang$^{1}$, Jungang Yang$^{1}$,\\ Wei An$^{1}$, Jingyi Yu$^{2}$, and Yulan Guo$^{1,3}$}
\institute{
$^{1}$National University of Defense Technology,\\
$^{2}$ShanghaiTech University,
$^{3}$Sun Yat-sen University.\\
\{wangyingqian16, yangjungang, yulan.guo\}@nudt.edu.cn}

\maketitle

\begin{abstract}
   Light field (LF) cameras record both intensity and directions of light rays, and capture scenes from a number of viewpoints. Both information within each perspective (i.e., spatial information) and among different perspectives (i.e., angular information) is beneficial to image super-resolution (SR). In this paper, we propose a spatial-angular interactive network (namely, LF-InterNet) for LF image SR. Specifically, spatial and angular features are first separately extracted from input LFs, and then repetitively interacted to progressively incorporate spatial and angular information. Finally, the interacted features are fused to super-resolve each sub-aperture image. Experimental results demonstrate the superiority of LF-InterNet over the state-of-the-art methods, i.e., our method can achieve high PSNR and SSIM scores with low computational cost, and recover faithful details in the reconstructed images.\footnote{Code is available at: https://github.com/YingqianWang/LF-InterNet.}.
\end{abstract}

\section{Introduction}
 Light field (LF) cameras provide multiple views of a scene, and thus enable many attractive applications such as post-capture refocusing \cite{wang2018selective}, depth sensing \cite{shin2018epinet}, saliency detection \cite{wang2019deep,zhang2019memory}, and de-occlusion \cite{DeOccNet}. However, LF cameras face a trade-off between spatial and angular resolution. That is, they either provide dense angular samplings with low image resolution (e.g., Lytro and RayTrix), or capture high-resolution (HR) sub-aperture images (SAIs) with sparse angular samplings (e.g., camera arrays \cite{wilburn2005high,venkataraman2013picam}). Consequently, many efforts have been made to improve the angular resolution through LF reconstruction \cite{wu2017light,wu2019learning,jin2020learning,shi2020learning}, or the spatial resolution through LF image super-resolution (SR) \cite{LFBM5D,resLF,GBSQ,LFNet,LFSSR,ATO}. In this paper, we focus on the LF image SR problem, namely, to reconstruct HR SAIs from their corresponding low-resolution (LR) SAIs.


 Image SR is a long-standing problem in computer vision. To achieve high reconstruction performance, SR methods need to incorporate as much useful information as possible from LR inputs. In the area of single image SR (SISR), good performance can be achieved by fully exploiting the neighborhood context (i.e., spatial information) in an image. Using the spatial information, SISR methods \cite{SRCNN2014,VDSR,EDSR,RCAN,SAN,SRGAN,ESRGAN} can successfully hallucinate missing details. In contrast, LFs record scenes from multiple views, and the complementary information among different views (i.e., angular information) can be used to further improve the performance of LF image SR.

 However, due to the complicated 4D structures of LFs, many LF image SR methods fail to fully exploit both the angular information and the spatial information, resulting in inferior SR performance. Specifically, in \cite{LFCNN2015,LFCNN2017,yuan2018light}, SAIs are first super-resolved separately using SISR methods \cite{SRCNN2014,EDSR}, and then fine-tuned together to incorporate the angular information. The angular information is ignored by these two-stage methods \cite{LFCNN2015,LFCNN2017,yuan2018light} during their upsampling process. In \cite{LFNet,resLF}, only part of SAIs are used to super-resolve one view, and the angular information in these discarded views is not incorporated. In contrast, Rossi et al. \cite{GBSQ} proposed a graph-based method to consider all angular views in an optimization process. However, this method \cite{GBSQ} cannot fully use the spatial information, and is inferior to recent deep learning-based SISR methods \cite{EDSR,RCAN,SAN}.

 Since spatial and angular information are highly coupled in 4D LFs and contribute to LF image SR in different manners, it is difficult for networks to perform well using these coupled information directly. In this paper, we propose a spatial-angular interactive network (i.e., LF-InterNet) to efficiently use spatial and angular information for LF image SR. Specifically, we design two convolutions (i.e., spatial/angular feature extractor) to extract and decouple spatial and angular features from input LFs. Then, we develop LF-InterNet to progressively interact the extracted features. Thanks to the proposed spatial-angular interaction mechanism, information in an LF can be effectively used in an efficient manner, and the SR performance is significantly improved. We perform extensive ablation studies to demonstrate the effectiveness of our model, and compare our method with state-of-the-art SISR and LF image SR methods from different perspectives, which demonstrate the superiority of our LF-InterNet.

\section{Related Works}
\vspace{-0.1cm}
\subsection{Single Image SR}
 In the area of SISR, deep learning-based methods have been extensively explored. Readers can refer to recent surveys \cite{wang2019deep,anwar2019deep,yang2019deep} for more details in SISR. Here, we only review several milestone works. Dong et al. \cite{SRCNN2014} proposed the first CNN-based SR method (i.e., SRCNN) by cascading 3 convolutional layers. Although SRCNN is shallow and simple, it achieves significant improvements over traditional SR methods \cite{timofte2013anchored,Jianchao2010Image,zeyde2010single}. Afterwards, SR networks became increasingly deep and complex, and thus more powerful in spatial information exploitation. Kim et al. \cite{VDSR} proposed a very deep SR network (i.e., VDSR) with 20 convolutional layers. Global residual learning is applied to VDSR to avoid slow convergence. Lim et al. \cite{EDSR} proposed an enhanced deep SR network (i.e., EDSR) and achieved substantial performance improvements by applying both local and global residual learning. Subsequently, Zhang et al. \cite{RDN} proposed a residual dense network (i.e., RDN) by combining residual connection and dense connection. RDN can fully extract hierarchical features for image SR, and thus achieve further improvements over EDSR. More recently, Zhang et al. \cite{RCAN} and Dai et al. \cite{SAN} further improved the performance of SISR by proposing residual channel attention network (i.e., RCAN) and second-order attention network (i.e., SAN). RCAN and SAN are the most powerful SISR methods to date and can achieve a very high reconstruction accuracy.

\vspace{-0.1cm}
\subsection{LF image SR}
 In the area of LF image SR, different paradigms have been proposed. Bishop et al. \cite{bishop2011light} first estimated the scene depth and then used a deconvolution approach to estimate HR SAIs. Wanner et al. \cite{wanner2013variational} proposed a variational LF image SR framework using the estimated disparity map. Farrugia et al. \cite{farrugia2017super} decomposed HR-LR patches into several subspaces, and achieved LF image SR via PCA analysis. Alain et al. \cite{LFBM5D} extended SR-BM3D \cite{BM3D} to LFs, and super-resolved SAIs using LFBM5D filtering. Rossi et al. \cite{GBSQ} formulated LF image SR as a graph optimization problem. These traditional methods \cite{bishop2011light,wanner2013variational,farrugia2017super,LFBM5D,GBSQ} use different approaches to exploit angular information, but perform inferior in spatial information exploitation as compared to recent deep learning-based methods.

 In the pioneering work of deep learning-based LF image SR (i.e., LFCNN \cite{LFCNN2015}), SAIs are super-resolved separately using SRCNN and fine-tuned in pairs to incorporate angular information. Similarly, Yuan et al. \cite{yuan2018light} proposed LF-DCNN, in which they used EDSR \cite{EDSR} to super-resolve each SAI and then fine-tuned the results. LFCNN and LF-DCNN handle the LF image SR problem in two stages and do not use angular information in the first stage. Wang et al. \cite{LFNet} proposed LFNet by extending BRCN \cite{BRCN} to LF image SR. In their method, SAIs from the same row or column are fed to a recurrent network to incorporate angular information. Zhang et al. \cite{resLF} stacked SAIs along different angular directions to generate input volumes, and then fed them to a multi-stream residual network named resLF. LFNet and resLF reduce 4D LF to 3D LF by using part of SAIs to super-resolve one view. Consequently, angular information in these discarded views cannot be incorporated. To consider all views for LF image SR, Yeung et al. \cite{LFSSR} proposed LFSSR to alternately shuffle LF features between SAI pattern and MacPI pattern for convolution. Jin et al. \cite{ATO} proposed an all-to-one LF image SR framework (i.e., LF-ATO) and performed structural consistency regularization to preserve the parallax structure among reconstructed views.

\vspace{-0.1cm}
\section{Method}
\vspace{-0.1cm}
\subsection{Spatial-Angular Feature Decoupling} \label{SAF Decoupling}
\begin{figure}[t]
\centering
\includegraphics[width=5.8cm]{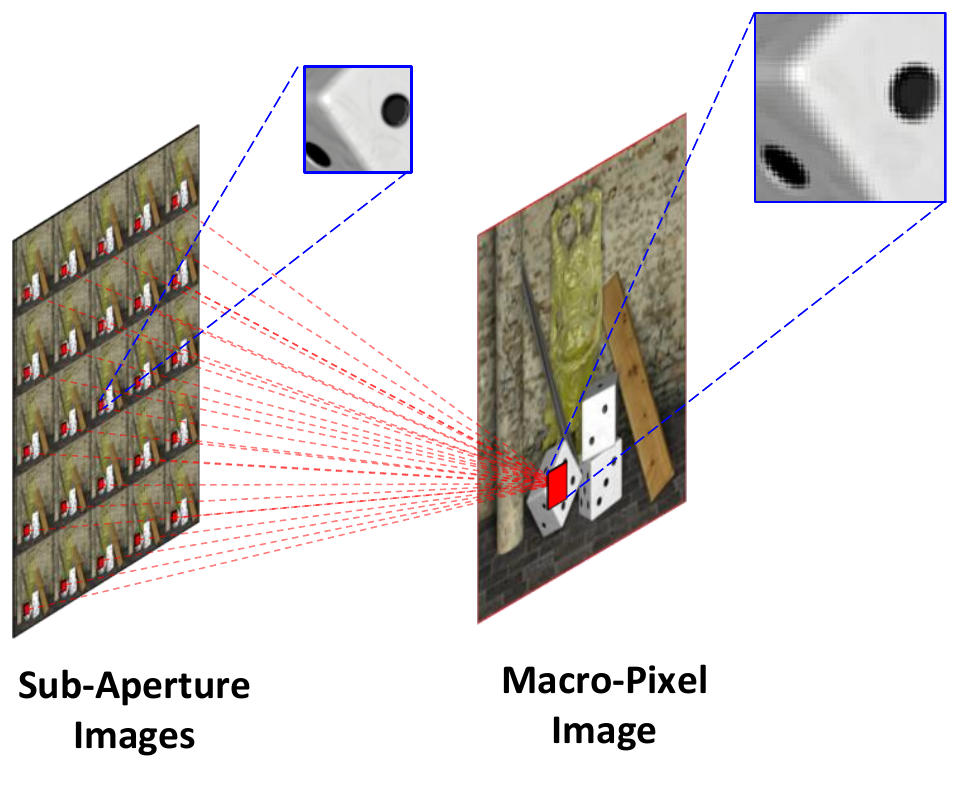}
\vspace{-0.2cm}
\caption{SAI array (left) and MacPI (right) representations of LFs. Both the SAI array and the MacPI representations have the same size of $\mathbb{R^{\mathrm{\mathit{UH}\times\mathit{VW}}}}$. Note that, to convert an SAI array representation into a MacPI representation, pixels at the same spatial coordinates of each SAI need to be extracted and organized according to their angular coordinates to generate a macro-pixel. Then, a MacPI can be generated by organizing these macro-pixels according to their spatial coordinates. More details are presented in the supplemental material.} \label{SAI_MPI}
\vspace{-0.3cm}
\end{figure}

 An LF has a 4D structure and can be denoted as $\mathcal{L}\in\mathbb{R^{\mathrm{\mathit{U}\times\mathit{V}\times\mathit{H}\times\mathit{W}}}}$, where $\mathit{U}$ and $\mathit{V}$ represent the angular dimensions (e.g., $\mathit{U}=3, \mathit{V}=4$ for a $3\times4$ LF), $\mathit{H}$ and $\mathit{W}$ represent the height and width of each SAI. Intuitively, an LF can be considered as a 2D angular collection of SAIs, and the SAI at each angular coordinate $(\mathit{u},\mathit{v})$ can be denoted as $\mathcal{L}\left(u,v,:,:\right)\in\mathbb{R^{\mathrm{\mathit{H}\times\mathit{W}}}}$. Similarly, an LF can also be organized into a 2D spatial collection of macro-pixels (namely, a MacPI). The macro-pixel at each spatial coordinate $(\mathit{h},\mathit{w})$ can be denoted as $\mathcal{L}\left(:,:,h,w\right)\in\mathbb{R^{\mathrm{\mathit{U}\times\mathit{V}}}}$. An illustration of these two LF representations is shown in Fig.~\ref{SAI_MPI}.

Since most methods use SAIs distributed in a square array as their input, we follow \cite{LFBM5D,GBSQ,LFCNN2015,LFCNN2017,LFSSR,resLF,ATO} to set $\mathit{U}=\mathit{V}=\mathit{A}$ in our method, where $\mathit{A}$ denotes the angular resolution. Given an LF of size $\mathbb{R^{\mathrm{\mathit{A}\times\mathit{A}\times\mathit{H}\times\mathit{W}}}}$, both a MacPI and an SAI array can be generated by organizing pixels according to corresponding patterns. Note that, when an LF is organized as an SAI array, the angular information is implicitly contained among different SAIs and thus is hard to extract. Therefore, we use the MacPI representation in our method and design spatial/angular feature extractors (SFE/AFE) to extract and decouple spatial/angular information.

\begin{figure}[t]
\centering
\includegraphics[width=7cm]{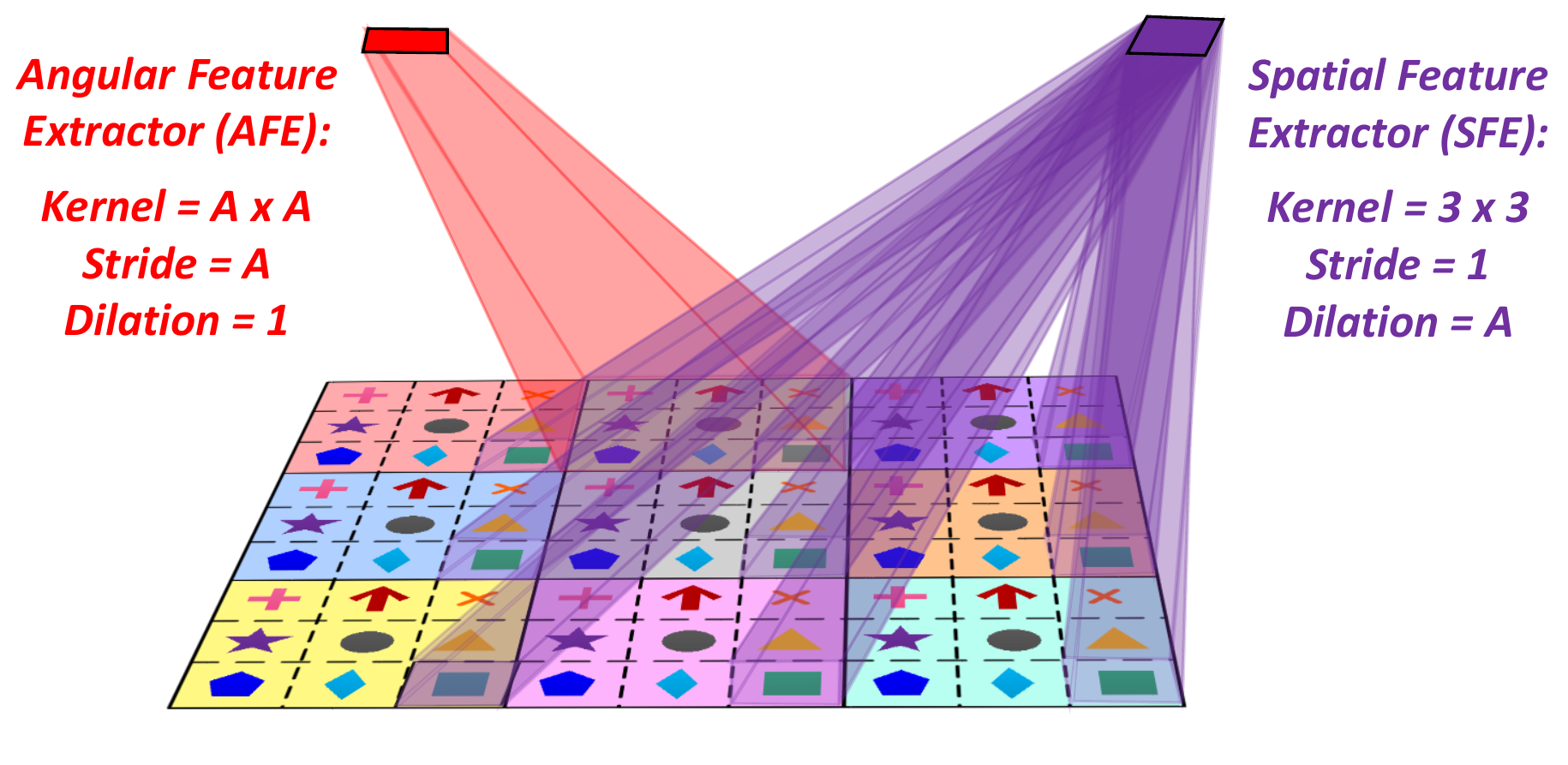}
\vspace{-0.2cm}
\caption{An illustration of angular and spatial feature extractors. Here, a LF of size $\mathbb{R^{\mathrm{3\times3\times3\times3}}}$ is used as a toy example. For better visualization, pixels from different SAIs are represented with different labels (e.g., red arrays or green squares), while different macro-pixels are paint with different background colors. Note that, AFE only extracts angular features and SFE only extracts spatial features, resulting in decoupling of  spatial-angular information.} \label{TwoConvs}
\vspace{-0.3cm}
\end{figure}

  Here, we use a toy example in Fig.~\ref{TwoConvs} to illustrate the angular and spatial feature extractors. Specifically, AFE is defined as a convolution with a kernel size of $\mathit{A}\times\mathit{A}$ and a stride of $\mathit{A}$. Padding is not performed so that features generated by AFE have a size of $\mathbb{R^{\mathrm{\mathit{H}\times\mathit{W}\times\mathit{C}}}}$, where $C$ represents the feature depth. In contrast, SFE is defined as a convolution with a kernel size of 3$\times$3, a stride of 1, and a dilation of $\mathit{A}$. We perform zero padding to ensure that the output features have the same spatial size $\mathit{AH}\times\mathit{AW}$ as the input MacPI. It is worth noting that, during angular feature extraction, each macro-pixel can be exactly convolved by AFE, while the information across different macro-pixels is not aliased. Similarly, during spatial feature extraction, pixels in each SAI can be convolved by SFE, while the angular information is not involved. In this way, the spatial and angular information in an LF is decoupled.

\begin{figure}[t]
\centering
\includegraphics[width=6.5cm]{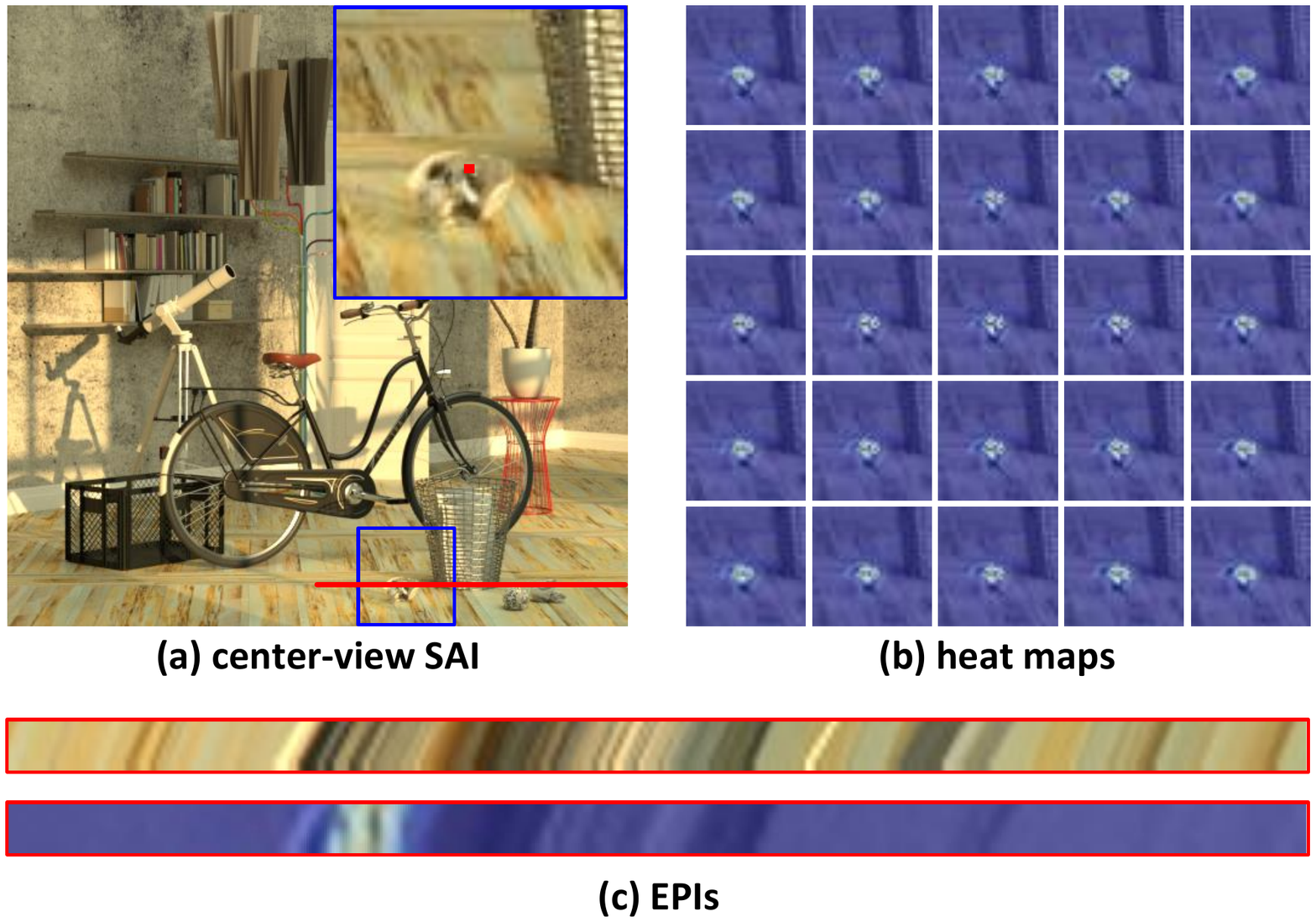}
\vspace{-0.2cm}
\caption{A visualization of the receptive field of our LF-InterNet. We performed 2$\times$SR on the 5$\times$5 center views of scene \textit{HCInew\_bicycle} \cite{HCInew}. (a) Center-view HR SAI. We select a target pixel (marked in red in the zoom-in region) at a shallow depth. (b) Highlighted input SAIs generated by the Grad-CAM method \cite{gradcam}.  A cluster of pixels in each SAI are highlighted as contributive pixels, which demonstrates the contribution of angular information. (c) Epipolar-plane images (EPIs) of the output LF (top) and the highlighted SAIs (bottom). It can be observed that the highlighted pixels in the bottom EPI have an enough receptive field to cover the slopes in the top EPI, which demonstates that our LF-InterNet can well handle the disparity problem in LF image SR.} \label{Heatmap}
\vspace{-0.3cm}
\end{figure}

 Due to the 3D property of real scenes, objects at different depths have different disparity values. Consequently, pixels of an object among different views cannot always locate at a single macro-pixel \cite{williem2018robust}. To address this problem, we enlarge the receptive field of our LF-InterNet by cascading multiple SFEs and AFEs in an interactive manner (see Fig.~\ref{Network}). Here, we use the Grad-CAM method \cite{gradcam} to visualize the receptive field of our LF-InterNet by highlighting contributive input regions. As shown in Fig.~\ref{Heatmap}, the angular information indeed contributes to LF image SR, and the receptive field is enough to cover the disparities in LFs.

\vspace{-0.1cm}
\subsection{Network Design} \label{Network Design}
\begin{figure}[t]
\centering
\includegraphics[width=12.0cm]{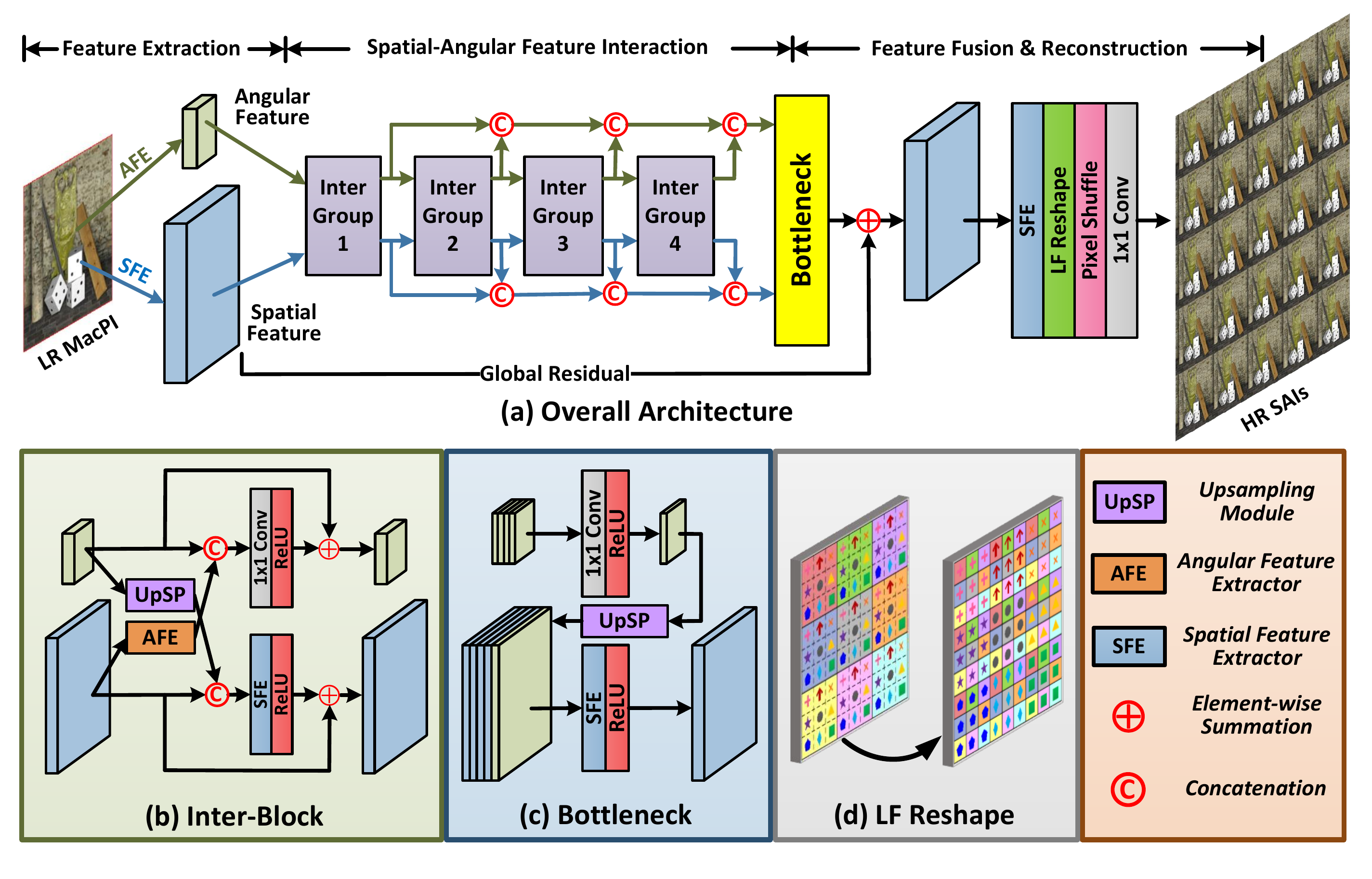}
\vspace{-0.3cm}
\caption{An overview of our LF-InterNet. Angular and spatial features are first extracted from the input MacPI, and then fed to a series of Inter-Groups (which consists of several cascaded Inter-Blocks) to achieve spatial-angular interaction. After LF reshape and pixel shuffling, HR SAIs are generated.} \label{Network}
\vspace{-0.5cm}
\end{figure}

 Our LF-InterNet takes an LR MacPI of size $\mathbb{R^{\mathrm{\mathit{AH}\times\mathit{AW}}}}$ as its input and produces an HR SAI array of size $\mathbb{R^{\mathrm{\mathit{\alpha AH}\times\mathit{\alpha AW}}}}$, where $\alpha$ denotes the upscaling factor. Following \cite{resLF,LFSSR,ATO}, we convert images into YCbCr color space, and only super-resolve the Y channel of images. An overview of our network is shown in Fig.~\ref{Network}.

\vspace{-0.3cm}
\subsubsection{Overall Architecture} \label{overall}
 Given an LR MacPI $\mathcal{I_{\mathit{LR}}}\in\mathbb{R^{\mathrm{\mathit{AH}\times\mathit{AW}}}}$, the angular and spatial features are first extracted by AFE and SFE, respectively.
\begin{equation}
\mathcal{F}_{A,0}=H_{A}\left(\mathcal{I_{\mathit{LR}}}\right),\quad
\mathcal{F}_{S,0}=H_{S}\left(\mathcal{I_{\mathit{LR}}}\right),
\end{equation}
 where $\mathcal{F}_{A,0}\in\mathbb{R^{\mathrm{\mathit{H}\times\mathit{W}\times\mathit{C}}}}$ and $\mathcal{F}_{S,0}\in\mathbb{R^{\mathrm{\mathit{AH}\times\mathit{AW}\times\mathit{C}}}}$ represent the extracted angular and spatial features, respectively. $H_{A}$ and $H_{S}$ represent the angular and spatial feature extractors (as described in Section~\ref{SAF Decoupling}), respectively. Once initial features are extracted, features $\mathcal{F}_{A,0}$ and $\mathcal{F}_{S,0}$ are further processed by a set of interaction groups (i.e., Inter-Groups) to achieve spatial-angular feature interaction:
\begin{equation}
\left(\mathcal{F}_{A,n},\mathcal{F}_{S,n}\right)=H_{IG,n}\left(\mathcal{F}_{A,n-1},\mathcal{F}_{S,n-1}\right), \quad \left(n=1,2,\cdots,N\right),
\end{equation}
 where $H_{IG,n}$ denotes the $\mathit{n^{th}}$ Inter-Group and $N$ denotes the total number of Inter-Groups.

 Inspired by RDN, we cascade all these Inter-Groups to fully use the information interacted at different stages. Specifically, features generated by each Inter-Group are concatenated and fed to a bottleneck block to fuse the interacted information. The feature generated by the bottleneck block is further added with the initial feature $\mathcal{F}_{S,0}$ to achieve global residual learning. The fused feature $\mathcal{F}_{S,t}$ can be obtained by
\begin{equation}
\mathcal{F}_{S,t}=H_{B}\left(\left[\mathcal{F}_{A,1},\cdots,\mathcal{F}_{A,N}\right], \left[\mathcal{F}_{S,1},\cdots,\mathcal{F}_{S,N}\right]\right)+\mathcal{F}_{S,0},
\end{equation}
 where $H_{B}$ denotes the bottleneck block, $\left[\cdot\right]$ denotes the concatenation operation. Finally, the fused feature $\mathcal{F}_{S,t}$ is fed to the reconstruction module, and an HR SAI array $\mathcal{I_{\mathit{SR}}}\in\mathbb{R^{\mathrm{\mathit{\alpha AH}\times\mathit{\alpha AW}}}}$ can be obtained by
\begin{equation}
\mathcal{I_{\mathit{SR}}}=H_{1\times1}\left(S_{pix}\left(R_{lf}\left(H_{S}\left(\mathcal{F}_{S,t}\right)\right)\right)\right),
\end{equation}
 where $R_{lf}$, $S_{pix}$, and $H_{1\times1}$ represent LF reshape, pixel shuffling, and $1\times1$ convolution, respectively.

\vspace{-0.3cm}
\subsubsection{Spatial-Angular Feature Interaction} \label{InterBlock}
 The basic module for spatial-angular interaction is the interaction block (i.e., Inter-Block). As shown in Fig.~\ref{Network} (b), the Inter-Block takes a pair of angular and spatial features as its inputs to achieve feature interaction. Specifically, the input angular feature is first upsampled by a factor of $A$. Since pixels in a MacPI can be unevenly distributed due to edges and occlusions in real scenes \cite{park2017robust}, we learn this discontinuity using a 1$\times$1 convolution and a pixel shuffling layer for angular-to-spatial upsampling. The upsampled angular feature is concatenated with the input spatial feature, and further fed to an SFE to incorporate the spatial and angular information. In this way, the complementary angular information can be used to guide spatial feature extraction. Simultaneously, the new angular feature is extracted from the input spatial feature by an AFE, and then concatenated with the input angular feature. The concatenated angular feature is further fed to a 1$\times$1 convolution to integrate and update the angular information. Note that, the fused angular and spatial features are added with their input features to achieve local residual learning. In this paper, we cascade $K$ Inter-Blocks in an Inter-Group, i.e., the output of an Inter-Block forms the input of its subsequent Inter-Block. In summary, the spatial-angular feature interaction can be formulated as
\begin{equation}
\mathcal{F}_{S,n}^{(k)}=H_{S}\left(\left[\mathcal{F}_{S,n}^{(k-1)},\left(\mathcal{F}_{A,n}^{(k-1)}\right)\uparrow\right]\right)+\mathcal{F}_{S,n}^{(k-1)},
\end{equation}
\begin{equation}
\mathcal{F}_{A,n}^{(k)}=H_{1\times1}\left(\left[\mathcal{F}_{A,n}^{(k-1)},H_{A}\left(\mathcal{F}_{S,n}^{(k-1)}\right)\right]\right)+\mathcal{F}_{A,n}^{(k-1)},
\left(k=1,2,\cdots,K\right),
\end{equation}
 where $\uparrow$ represents the upsampling operation, $\mathcal{F}_{S,n}^{(k)}$ and $\mathcal{F}_{A,n}^{(k)}$ represent the output spatial and angular features of the $k^{th}$ Inter-Block in the $n^{th}$ Inter-Group, respectively.

\vspace{-0.3cm}
\subsubsection{Feature Fusion and Reconstruction} \label{Fusion}
 The objective of this stage is to fuse the interacted features to reconstruct an HR SAI array. The fusion and reconstruction stage mainly consists of bottleneck fusion (as shown in Fig.~\ref{Network} (c)), LF reshape (as shown in Fig.~\ref{Network} (d)), pixel shuffling, and final reconstruction.

 In the bottleneck, the concatenated angular features $\left[\mathcal{F}_{A,1},\cdots,\mathcal{F}_{A,N}\right]\in\mathbb{R^{\mathrm{\mathit{H}\times\mathit{W}\times\mathit{NC}}}}$ are first fed to a 1$\times$1 convolution and a ReLU layer to generate a feature map $\mathcal{F}_{A}\in\mathbb{R^{\mathrm{\mathit{H}\times\mathit{W}\times\mathit{C}}}}$. Then, the squeezed angular feature $\mathcal{F}_{A}$ is upsampled and concatenated with  spatial features. The final fused feature $\mathcal{F}_{S,t}$ can be obtained as
\begin{equation}
\mathcal{F}_{S,t}=H_{S}\left(\left[\mathcal{F}_{S,1},\cdots,\mathcal{F}_{S,N},\left(\mathcal{F}_{A}\right)\uparrow\right]\right)+\mathcal{F}_{S,0}.
\end{equation}

 After feature fusion, we apply another SFE layer to extend the channel size of $\mathcal{F}_{S,t}$ to $\alpha^{2}C$ for pixel shuffling \cite{PixelShuffle}. However, since $\mathcal{F}_{S,t}$ is organized in the MacPI pattern, we apply LF reshape to convert $\mathcal{F}_{S,t}$ into a SAI array representation for pixel shuffling. To achieve LF reshape, we first extract pixels with the same angular coordinates in the MacPI feature, and then re-organize these pixels according to their spatial coordinates, which can be formulated as
\begin{equation}
\mathcal{I_{\mathit{SAIs}}}\left(x,y\right)=\mathcal{I_{\mathit{MacPI}}}\left(\xi,\eta\right),
\end{equation}
where
\begin{equation}
x=H\left(\xi-1\right)+\lfloor \xi/A\rfloor\left(1-AH\right)+1,
\end{equation}
\begin{equation}
y=W\left(\eta-1\right)+\lfloor \eta/A\rfloor\left(1-AW\right)+1.
\end{equation}

 Here, $x=1,2,\cdots,AH$ and $y=1,2,\cdots,AW$ denote the pixel coordinates in the output SAI arrays, $\xi$ and $\eta$ denote the corresponding coordinates in the input MacPI, $\lfloor\cdot\rfloor$ represents the round-down operation. The derivation of Eqs.~(9) and (10) is presented in the supplemental material. Finally, a 1$\times$1 convolution is applied to squeeze the number of feature channels to 1 for HR SAI reconstruction.
\vspace{-0.5cm}
\section{Experiments}
\vspace{-0.1cm}
 In this section, we first introduce the datasets and our implementation details. Then we conduct ablation studies to investigate our network. Finally, we compare our LF-InterNet to several state-of-the-art LF image SR and SISR methods.

\vspace{-0.3cm}
\subsection{Datasets and Implementation Details}
 As listed in Table~\ref{Datasets}, we used 6 public LF datasets \cite{EPFL,HCInew,HCIold,INRIA,STFgantry,STFlytro} in our experiments. All the LFs in the training and test sets have an angular resolution of 9$\times$9. In the training stage, we cropped each SAI into patches of size 64$\times$64, and then followed the existing SR methods \cite{VDSR,EDSR,RCAN,SAN,LFSSR,resLF} to use bicubic downsampling with a factor of $\alpha \left(\alpha=2,4\right)$ to generate LR patches. The generated LR patches were re-organized into a MacPI pattern to form the input of our network. The $L_1$ loss function was used since it can generate good results for the SR task and is robust to outliers \cite{SRlibrary}. Following \cite{resLF}, we augmented the training data by 8 times using random flipping and 90-degree rotation. Note that, during each data augmentation, all SAIs need to be flipped and rotated along both spatial and angular directions to maintain their LF structures.

\begin{table}[t]
\centering
\scriptsize
\caption{Datasets used in our experiments.}\label{Datasets}
\begin{tabular}{|l|c|c|c|c|c|c|c|}
\hline
&~EPFL \cite{EPFL}~&~HCInew \cite{HCInew}~&~HCIold \cite{HCIold}~&~INRIA \cite{INRIA}~&~STFgantry \cite{STFgantry}~&~STFlytro \cite{STFlytro}\\
\hline
Training ~& 70 & 20 & 10 & 35 & 9 & 250 \\
Test  & 10 & 4  & 2 & 5 & 2 & 50\\
\hline
\end{tabular}
\vspace{-0.3cm}
\end{table}
 By default, we used the model with $N=4$, $K=4$, $C=64$, and angular resolution of 5$\times$5 for both 2$\times$ and 4$\times$SR. We also investigated the performance of other branches of our LF-InterNet in Section~\ref{Ablation}. We used PSNR and SSIM as quantitative metrics for performance evaluation. Note that, PSNR and SSIM were separately calculated on the Y channel of each SAI. To obtain the overall metric score for a dataset with $M$ scenes (each with an angular resolution of $A\times A$), we first obtain the score for a scene by averaging its $A^{2}$ scores, and then obtain the overall score by averaging the scores of all $M$ scenes.

 Our LF-InterNet was implemented in PyTorch on a PC with an Nvidia RTX 2080Ti GPU. Our model was initialized using the Xavier method \cite{Xavier} and optimized using the Adam method \cite{Adam}. The batch size was set to 12 and the learning rate was initially set to 5$\times10^{-4}$ and decreased by a factor of 0.5 for every 10 epochs. The training was stopped after 40 epochs and took about one day.

\vspace{-0.3cm}
\subsection{Ablation Study} \label{Ablation}
In this subsection, we compare the performance of our LF-InterNet with different architectures and angular resolutions to investigate the potential benefits introduced by different design choices.

\textbf{Angular Information.} We investigated the benefit of angular information by removing the angular path in LF-InterNet. That is, we only use SFE for LF image SR. Consequently, the network is identical to a SISR network, and can only incorporate spatial information within each SAI. As shown in Table~\ref{AblationModel}, only using the spatial information, the network (i.e., LF-InterNet-SpatialOnly) achieves a PSNR of 29.98 and a SSIM of 0.897, which are significantly inferior to LF-InterNet. Therefore, the benefit of angular information to LF image SR is clearly demonstrated.

\textbf{Spatial Information.} To investigate the benefit introduced by spatial information, we changed the kernel size of all SFEs from 3$\times$3 to 1$\times$1. In this case, the spatial information cannot be exploited and integrated by convolutions. As shown in Table~\ref{AblationModel}, the performance of LF-InterNet-AngularOnly is even inferior to bicubic interpolation. That is because, neighborhood context in an image is highly significant in recovering details. It is clear that spatial information plays a major role in LF image SR, while angular information can only be used as a complementary part to spatial information but cannot be used alone.

\textbf{Information Decoupling.} To investigate the benefit of spatial-angular information decoupling, we stacked all SAIs along the channel dimension as input, and used 3$\times$3 convolutions with a stride of 1 to extract both spatial and angular information from these stacked images. Note that, the cascaded framework with global and local residual learning was maintained to keep the overall network architecture unchanged. To achieve fair comparison, we adjusted the feature depths to keep the model size (i.e., LF-InterNet-SAcoupled\_1) or computational complexity (i.e., LF-InterNet-SAcoupled\_2) comparable to LF-InterNet. As shown in Table~\ref{AblationModel}, both LF-InterNet-SAcoupled\_1 and LF-InterNet-SAcoupled\_2 are inferior to LF-InterNet. It is clearly demonstrated that, our LF-InterNet can handle the 4D LF structure and achieve LF image SR much more efficiently by using the proposed spatial-angular feature decoupling mechanism.
\begin{table}[t]
\vspace{-0.3cm}
\caption{Comparative results achieved on the STFlytro dataset \cite{STFlytro} by several variants of our LF-InterNet for 4$\times$SR. Note that, we carefully adjusted the feature depths of different variants to make their model size comparable. FLOPs are computed with an input MacPI of size 160$\times$160. The results of bicubic interpolation are listed as baselines.}\label{AblationModel}
\centering
\scriptsize
\begin{tabular}{|l|c|c|c|c|}
\hline
Model~~ & ~~PSNR~~ & ~~SSIM~~ & ~~Params.~~ & ~~FLOPs~~\\
\hline
Bicubic                 & 27.84 & 0.855 & ---  & ---   \\
LF-InterNet-SpatialOnly~~ & 29.98 & 0.897 & 5.40M & 134.7G \\
LF-InterNet-AngularOnly~~ & 26.57 & 0.823 & 5.43M & 13.4G\\
LF-InterNet-SAcoupled\_1~~   & 31.11 & 0.918 & 5.42M & 5.46G\\
LF-InterNet-SAcoupled\_2~~   & 31.17 & 0.919 & 50.8M & 50.5G\\
LF-InterNet             & \textbf{31.65} & \textbf{0.925} & 5.23M & 50.1G\\
\hline
\end{tabular}
\vspace{-0.3cm}
\end{table}

\begin{table}[b]
 \vspace{-0.8cm}
 \caption{Comparative results achieved on the STFlytro dataset \cite{STFlytro} by our LF-InterNet with different number of interactions for 4$\times$SR.}\label{AblationInteraction}
 \centering
\scriptsize
 \begin{tabular}{|cccc|c|c|c|}
 \hline
 ~~IG\_1~~ & ~~IG\_2~~ &~~IG\_3~~ &~~IG\_4~~ & ~~PSNR~~ & ~~SSIM~~ \\
 \hline
              &              &              &              & 29.84 & 0.894  \\
 \hline
 $\checkmark$ &              &              &              & 31.44 & 0.922\\
 \hline
 $\checkmark$ & $\checkmark$ &              &              & 31.61 & 0.924\\
 \hline
 $\checkmark$ & $\checkmark$ & $\checkmark$ &              & 31.66 & 0.925\\
 \hline
 $\checkmark$ & $\checkmark$ & $\checkmark$ & $\checkmark$ & \textbf{31.84} & \textbf{0.927} \\
 \hline
 \end{tabular}
 \end{table}

\textbf{Spatial-Angular Interaction.} We investigated the benefits introduced by our spatial-angular interaction mechanism. Specifically, we canceled feature interaction in each Inter-Group by removing upsampling and AFE modules in each Inter-Block (see Fig.~\ref{Network} (b)). In this case, spatial and angular features can only be processed separately. When all interactions are removed, these spatial and angular features can only be incorporated by the bottleneck block. Table~\ref{AblationInteraction} presents the results achieved by our LF-InterNet with different numbers of interactions. It can be observed that, without any feature interaction, our network achieves a very low reconstruction accuracy (i.e., 29.84 in PSNR and 0.894 in SSIM). That is because, the angular and spatial information cannot be effectively incorporated by the bottleneck block without feature interactions. As the number of interactions increases, the performance is steadily improved. This clearly demonstrates the effectiveness of our spatial-angular feature interaction mechanism.

\begin{table}[t]
\centering
\scriptsize
\vspace{-0.2cm}
\caption{Comparisons of different approaches for angular-to-spatial upsampling.}\label{pixSF}
\begin{tabular}{|l|ccc|ccc|}
\hline
Model ~~&~~ Scale ~~&~~  PSNR~~&~~ SSIM~~&~~ Scale~~&~~ PSNR~~&~~ SSIM~~ \\
\hline
LF-InterNet-nearest ~~&~~ 2$\times$ ~~&~~  38.60 ~~&~~  0.982 ~~&~~  4$\times$ ~~&~~  31.65 ~~&~~  0.925  \\
LF-InterNet-bilinear ~~&~~ 2$\times$ ~~&~~  37.67 ~~&~~  0.976 ~~&~~   4$\times$ ~~&~~  30.71 ~~&~~  0.911  \\
LF-InterNet ~~&~~ 2$\times$ ~~&~~  \textbf{38.81} ~~&~~  \textbf{0.983}  ~~&~~  4$\times$ ~~&~~  \textbf{31.84} ~~&~~  \textbf{0.927}  \\
\hline
\end{tabular}
\vspace{-0.2cm}
\end{table}

\textbf{Angular-to-Spatial Upsampling.} To demonstrate the effectiveness of the pixel shuffling layer used in angular-to-spatial upsampling, we introduced two variants by replacing pixel shuffling with nearest upsampling and bilinear upsampling, respectively. It can be observed from Table \ref{pixSF} that LF-InterNet-bilinear achieves much lower PSNR and SSIM scores than LF-InterNet-nearest and LF-InterNet. That is because, bilinear interpolation  introduces aliasing among macro-pixels during angular-to-spatial upsampling, resulting in ambiguities in spatial-angular feature decoupling and interaction. In contrast, both nearest upsampling and pixel shuffling do not introduce aliasing and thus achieve improved performance. Moreover, since pixels in a macro-pixel can be unevenly distributed due to edges and occlusions in real scenes \cite{park2017robust}, pixel shuffling achieves a further improvement over nearest upsampling due to its discontinuity modeling within macro-pixels.

\begin{table}[t]
 \vspace{-0.1cm}
\caption{Comparative results achieved on the STFlytro dataset \cite{STFlytro} by our LF-InterNet with different angular resolutions for 2$\times$ and 4$\times$SR.}\label{AblationAngRes}
\centering
\scriptsize
\begin{tabular}{|c|ccc|ccc|}
\hline
~~AngRes~~ & ~~Scale~~ & ~~PSNR~~ & ~~SSIM~~ & ~~Scale~~ & ~~PSNR~~ & ~~SSIM~~ \\[0.5pt]
\hline
3$\times$3 & 2$\times$ & 37.95 & 0.980 & 4$\times$ & 31.30 & 0.918 \\
5$\times$5 & 2$\times$ & 38.81 & 0.983 & 4$\times$ & 31.84 & 0.927  \\
7$\times$7 & 2$\times$  & 39.05 & 0.984 & 4$\times$  & 32.04 & 0.931  \\
9$\times$9 & 2$\times$  & \textbf{39.08} & \textbf{0.985} & 4$\times$  & \textbf{32.07} & \textbf{0.933} \\
\hline
\end{tabular}
 \vspace{-0.3cm}
\end{table}

\textbf{Angular Resolution.} In this section, we analyze the performance of LF-InterNet with different angular resolutions. Specifically, we extracted the central SAIs with size of $A\times A \left(A=3,5,7,9\right)$ from the input LFs, and trained different models for both 2$\times$ and 4$\times$SR. As shown in Table~\ref{AblationAngRes}, the PSNR and SSIM values for both 2$\times$ and 4$\times$ SR are improved as the angular resolution is increased. That is because, additional views provide rich angular information for LF image SR. It is also notable that, the improvement tends to be saturated when the angular resolution is further increased from 7$\times$7 to 9$\times$9 (with only 0.03 dB improvement in PSNR). That is because, the complementary information provided by additional views is already sufficient. Since the angular information has been fully exploited for an angular resolution of 7$\times$7, a further increase of views can only provide minor performance improvement.

\vspace{-0.3cm}
\subsection{Comparison to the State-of-the-arts}

  We compare our method to 6 SISR methods \cite{VDSR,EDSR,RCAN,SAN,SRGAN,ESRGAN} and 5 LF image SR methods \cite{LFBM5D,GBSQ,LFSSR,resLF,ATO}. Bicubic interpolation was used as baselines.

\begin{table}[t]
\vspace{-0.2cm}
\caption{PSNR/SSIM values achieved by different methods for 2$\times$ and 4$\times$SR. The best results are in \textcolor{red}{red} and the second best results are in \textcolor{blue}{blue}.}\label{Comparison}
\centering
\scriptsize
\renewcommand\arraystretch{1}
\vspace{0.1cm}
\begin{tabular}{|l|c|c|c|c|c|c|c|}
\hline
Method & Scale & EPFL & HCInew & HCIold & INRIA & STFgantry & STFlytro\\
\hline
Bicubic  &$2\times$                             & 29.50/0.935 & 31.69/0.934 & 37.46/0.978 & 31.10/0.956 & 30.82/0.947 & 33.02/0.950\\
VDSR \cite{VDSR} &$2\times$           & 32.01/0.959 & 34.37/0.956 & 40.34/0.985 & 33.80/0.972 & 35.80/0.980 & 35.91/0.970\\
EDSR \cite{EDSR}  &$2\times$           & 32.86/0.965 & 35.02/0.961 & 41.11/0.988 & 34.61/0.977 & 37.08/0.985 & 36.87/0.975 \\
RCAN  \cite{RCAN}  &$2\times$        & 33.46/0.967 & 35.56/0.963 & 41.59/0.989 & 35.18/0.978 & 38.18/0.988 & 37.32/0.977\\
SAN  \cite{SAN}     &$2\times$           & 33.36/0.967  & 35.51/0.963  & 41.47/0.989  & 35.15/0.978  & 37.98/0.987  & 37.26/0.976 \\
LFBM5D \cite{LFBM5D} &$2\times$  & 31.15/0.955 & 33.72/0.955 & 39.62/0.985 & 32.85/0.969 & 33.55/0.972 & 35.01/0.966\\
GB \cite{GBSQ}  &$2\times$         & 31.22/0.959 & 35.25/0.969 & 40.21/0.988 & 32.76/0.972 & 35.44/0.983 & 35.04/0.956 \\
resLF \cite{resLF} &$2\times$             & 33.22/0.969 & 35.79/0.969 & 42.30/0.991 & 34.86/0.979 & 36.28/0.985 & 35.80/0.970\\
LFSSR \cite{LFSSR} &$2\times$        & 34.15/\textcolor{blue}{0.973} & 36.98/0.974 & 43.29/0.993 & 35.76/\textcolor{blue}{0.982} & 37.67/0.989 & 37.57/0.978\\
LF-ATO \cite{ATO} &$2\times$               & \textcolor{blue}{34.49}\textcolor{red}{/0.976} & \textcolor{red}{37.28/0.977} & \textcolor{blue}{43.76/0.994} & \textcolor{blue}{36.21}/\textcolor{red}{0.984} & \textcolor{red}{39.06/0.992} & \textcolor{blue}{38.27/0.982} \\
LF-InterNet  &$2\times$                       & \textcolor{red}{34.76/0.976} & \textcolor{blue}{37.20/0.976} & \textcolor{red}{44.65/0.995} & \textcolor{red}{36.64/0.984} & \textcolor{blue}{38.48/0.991} & \textcolor{red}{38.81/0.983}\\
\hline
Bicubic    &$4\times$                           & 25.14/0.831 & 27.61/0.851 & 32.42/0.934 & 26.82/0.886 & 25.93/0.843 & 27.84/0.855 \\
VDSR \cite{VDSR}   &$4\times$        & 26.82/0.869 & 29.12/0.876 & 34.01/0.943 & 28.87/0.914 & 28.31/0.893 & 29.17/0.880  \\
EDSR \cite{EDSR}  &$4\times$          & 27.82/0.892 & 29.94/0.893 & 35.53/0.957 & 29.86/0.931 & 29.43/0.921 & 30.29/0.903 \\
RCAN \cite{RCAN}   &$4\times$       & 28.31/0.899 & 30.25/0.896 & 35.89/0.959 & 30.36/0.936 & 30.25/ 0.934 & 30.66/0.909\\
SAN \cite{SAN}      &$4\times$          & 28.30/0.899 & 30.25/0.898 & 35.88/0.960 & 30.29/0.936 & 30.30/0.933 & 30.71/0.909  \\
SRGAN \cite{SRGAN} &$4\times$    &  26.85/0.870 & 28.95/0.873 & 34.03/0.942 & 28.85/0.916 & 28.19/0.898 & 29.28/0.883\\
ESRGAN \cite{ESRGAN}&$4\times$ &  25.59/0.836 & 26.96/0.819 & 33.53/0.933 & 27.54/0.880 & 28.00/0.905 & 27.09/0.826 \\
LFBM5D \cite{LFBM5D} &$4\times$ & 26.61/0.869 & 29.13/0.882 & 34.23/0.951 & 28.49/0.914 & 28.30/0.900 & 29.07/0.881\\
GB \cite{GBSQ}   &$4\times$       & 26.02/0.863 & 28.92/0.884 & 33.74/0.950 & 27.73/0.909 & 28.11/0.901 & 28.37/0.873  \\
resLF \cite{resLF} &$4\times$            & 27.86/0.899 & 30.37/0.907 & 36.12/0.966 & 29.72/0.936 & 29.64/0.927 & 28.94/0.891  \\
LFSSR \cite{LFSSR}&$4\times$        & \textcolor{blue}{29.16/0.915} & 30.88/0.913 & \textcolor{blue}{36.90}/0.970 & 31.03/0.944 & 30.14/0.937 & \textcolor{blue}{31.21/0.919} \\
LF-ATO \cite{ATO}&$4\times$               & \textcolor{blue}{29.16}/\textcolor{red}{0.917} & \textcolor{red}{31.08/0.917} & \textcolor{red}{37.23}/\textcolor{blue}{0.971} & \textcolor{blue}{31.21}/\textcolor{red}{0.950} & \textcolor{red}{30.78/0.944} & 30.98/0.918 \\
LF-InterNet  &$4\times$                     & \textcolor{red}{29.52/0.917} & \textcolor{blue}{31.01}/\textcolor{red}{0.917} & \textcolor{red}{37.23/0.972} & \textcolor{red}{31.65/0.950} & \textcolor{blue}{30.44/0.941} & \textcolor{red}{31.84/0.927} \\
\hline
\end{tabular}
\vspace{-0.2cm}
\end{table}

\textbf{Quantitative Results.} Quantitative results in Table~\ref{Comparison} demonstrate the state-of-the-art performance of our LF-InterNet on all the 6 test datasets. Thanks to the use of angular information, our method achieves an improvement of 1.54 dB (2$\times$SR) and 1.00 dB (4$\times$SR) in PSNR over the powerful SISR method RCAN \cite{RCAN}. Moreover, our LF-InterNet can achieve a comparable PSNR and SSIM scores as compared to the most recent LF image SR method LF-ATO \cite{ATO}.

\begin{figure}[t]
\centering
\vspace{-0.1cm}
\includegraphics[width=12.2cm]{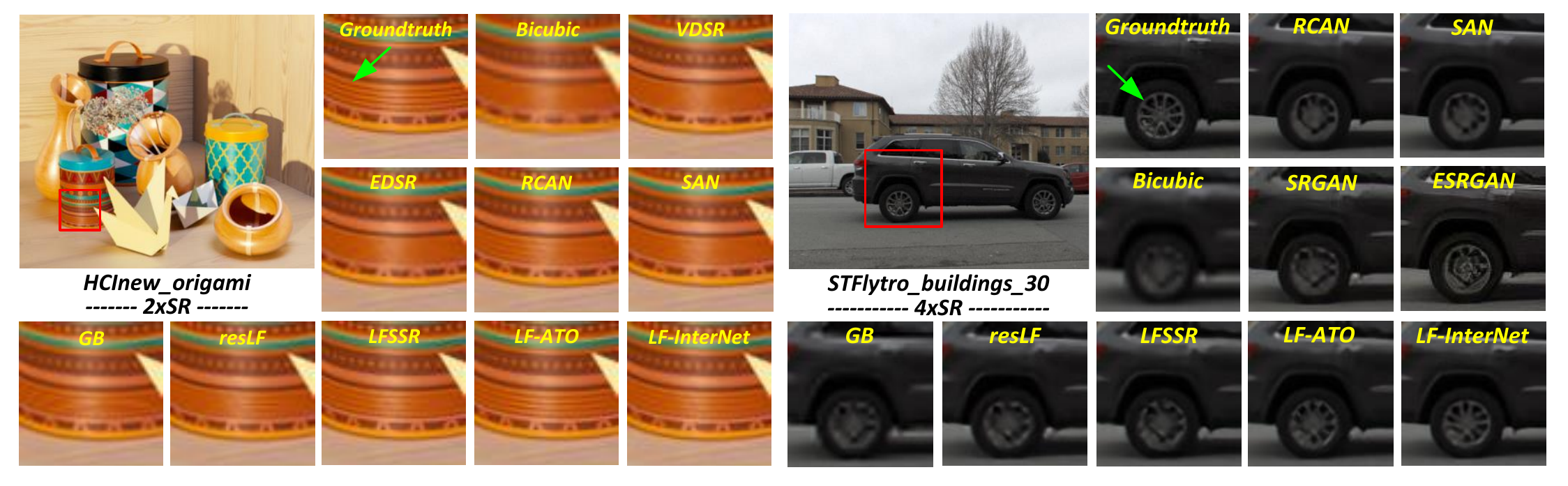}
\vspace{-0.4cm}
\caption{Visual results of 2$\times$/4$\times$SR.} \label{VisualResults}
\vspace{-0.5cm}
\end{figure}

\textbf{Qualitative Results.} Qualitative results of $2\times$/$4\times$SR are shown in Fig.~\ref{VisualResults}, with more visual comparisons being provided in our supplemental material. Our LF-InterNet can well preserve the textures and details (e.g., the horizontal stripes in the scene \textit{HCInew\_origami}) in the super-resolved images. In contrast, state-of-the-art SISR methods RCAN \cite{RCAN} and SAN \cite{SAN} produce oversmoothed images with poor details. The visual superiority of our method is more obvious for 4$\times$SR. That is because, the input LR images are severely degraded by the down-sampling operation, and the process of 4$\times$SR is highly ill-posed. In such cases, some perceptual-oriented methods (e.g., SRGAN \cite{SRGAN} and ESRGAN \cite{ESRGAN}) use spatial information only to hallucinate missing details, resulting in ambiguous and even fake textures (e.g., wheel in scene \textit{STFlytro\_buildings}). In contrast, our method can use complementary angular information among different views to produce more faithful results.

\begin{table}[t]
\centering
\scriptsize
\vspace{-0.1cm}
\caption{Comparisons of the number of parameters (\#Params.) and FLOPs for 2$\times$ and 4$\times$SR. Note that, the FLOPs is calculated on an input LF with a size of 5$\times$5$\times$32$\times$32, and the PSNR and SSIM scores are averaged over the 6 test datasets \cite{EPFL,HCInew,HCIold,INRIA,STFgantry,STFlytro} in Table \ref{Comparison}.}\label{tabEfficiency}
\begin{tabular}{|l|c|c|c|c|c|c|c|c|}
\hline
Method & Scale &  \#Params. & FLOPs(G) & PSNR/SSIM & Scale &  \#Params.  & FLOPs(G) & PSNR/SSIM \\
\hline
RCAN \cite{RCAN} & 2$\times$ & 15.44M  & 15.71$\times$25 &  36.88/0.977 & 4$\times$& 15.59M  & 16.34$\times$25 & 30.95/0.922\\
SAN \cite{SAN} & 2$\times$ & 15.71M  & 16.05$\times$25  & 36.79/0.977 & 4$\times$& 15.86M & 16.67$\times$25 &  31.96/0.923\\
resLF \cite{resLF} & 2$\times$ & 6.35M & 37.06 &  36.38/0.977 & 4$\times$ &  6.79M & 39.70  &  30.08/0.916\\
LFSSR \cite{LFSSR} & 2$\times$ & 0.81M & 25.70 & 37.57/0.982  & 4$\times$ &  1.61M & 128.44  & 31.55/0.933 \\
LF-ATO \cite{ATO} & 2$\times$ & 1.51M & 597.66 & 38.18/0.984  & 4$\times$ &  1.66M & 686.99  & 31.74/0.937 \\
LF-InterNet\_32 ~& 2$\times$ &  1.20M & 11.87 &  37.88/0.983 & 4$\times$ & 1.31M & 12.53 &  31.57/0.933\\
LF-InterNet\_64 ~& 2$\times$ &  4.80M & 47.46 &  38.42/0.984 & 4$\times$ &  5.23M & 50.10 &  31.95/0.937\\
\hline
\end{tabular}
\vspace{-0.4cm}
\end{table}

\textbf{Efficiency.} We compare our LF-InterNet to several competitive methods \cite{RCAN,SAN,resLF,LFSSR,ATO} in terms of  the number of parameters and FLOPs. As shown in Table \ref{tabEfficiency}, our LF-InterNet achieves superior SR performance with reasonable number of parameters and FLOPs. Note that, although LF-ATO has very small model sizes (i.e., 1.51M for $2\times$SR and 1.66M for $4\times$SR), its FLOPs are very high since it uses the \textit{All-to-One} strategy to separately super-resolve individual views in a sequence. In contrast, our method (i.e., LF-InterNet\_64) super-resolves all views simultaneously, and achieves a comparable or even better performance than LF-ATO with significantly lower FLOPs. It is worth noting that, even the feature depth of our model is halved to 32, our method (i.e., LF-InterNet\_32) can still achieve promising PSNR$/$SSIM scores, which are comparable to LFSSR and higher than RCAN, SAN, and resLF. The above comparisons clearly demonstrate the high efficiency of our network architecture.

\begin{figure}[t]
\centering
\vspace{-0.0cm}
\includegraphics[width=12.2cm]{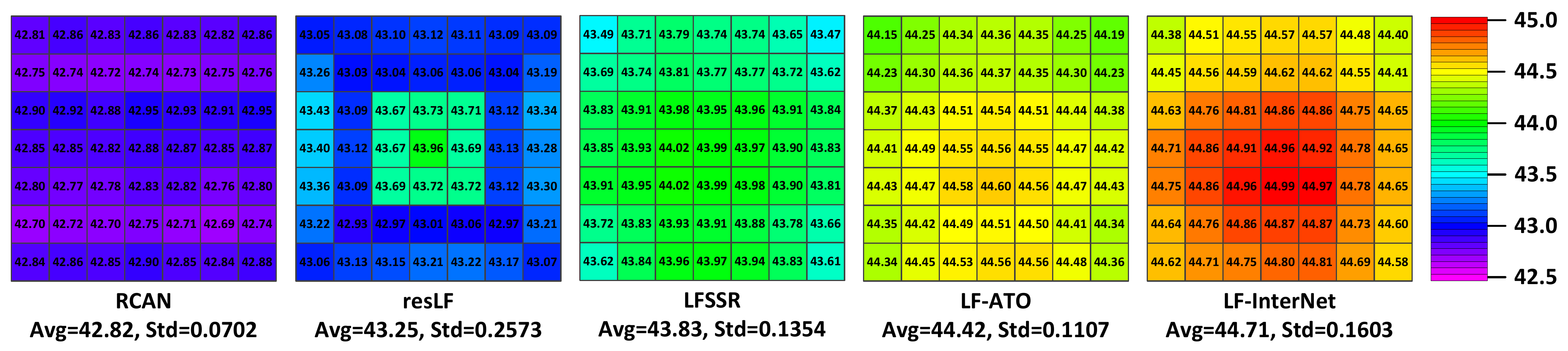}
\vspace{-0.4cm}
\caption{Comparative results (i.e., PSNR values) achieved on each perspective of scene \textit{HCIold\_MonasRoom}. Here, 7$\times$7 input views are used to perform 2$\times$SR. We use standard deviation (Std) to represent their uniformity. Our LF-InterNet achieves high reconstruction quality with a relatively balanced distribution.} \label{PwrtP}
\vspace{-0.3cm}
\end{figure}
\textbf{Performance w.r.t. Perspectives.} Since our LF-InterNet can super-resolve all SAIs in an LF, we further investigate the reconstruction quality with respect to different perspectives. We followed \cite{resLF} to use the central 7$\times$7 views of scene \textit{HCIold\_MonasRoom} to perform 2$\times$SR, and used PSNR for performance evaluation.  Note that, due to the changing perspectives, the contents of different SAIs are not identical, resulting in inherent PSNR variations. Therefore, we evaluate this variation by using RCAN to perform SISR on each SAI. Results are reported and visualized in Fig.~\ref{PwrtP}. Since resLF uses part of views to super-resolve different perspectives, the reconstruction qualities of resLF for non-central views are relatively low. In contrast, LFSSR, LF-ATO and our LF-InterNet can use the angular information from all input views to super-resolve each view, and thus achieve a relatively balanced distribution (i.e., lower Std scores) among different perspectives. The reconstruction quality (i.e., PSNR scores) of LF-InterNet is slightly higher than those of LFSSR and LF-ATO on this scene.

\begin{table}
\vspace{-0.2cm}
\caption{Comparative results achieved on the UCSD dataset for 2$\times$ and 4$\times$SR.}\label{tabGeneralization}
\centering
\scriptsize
\begin{tabular}{|l|cc|cc|}
\hline
Method ~~ & ~~ Scale ~~ & ~~ PSNR$/$SSIM ~~ & ~~ Scale ~~ & ~~ PSNR$/$SSIM \\
\hline
RCAN \cite{RCAN}~ & ~~ 2$\times$ ~~ & ~~ 41.63/0.983 ~~ & ~~ 4$\times$ ~~ & ~~ 36.49/0.955 \\
SAN \cite{SAN}~ & ~~ 2$\times$ ~~ & ~~ 41.56/0.983 ~~ & ~~ 4$\times$ ~~ & ~~ 36.57/0.956  \\
resLF \cite{resLF}~ & ~~ 2$\times$  ~~ & ~~ 41.29/0.982 ~~ & ~~ 4$\times$  ~~ & ~~ 35.89/0.953  \\
LFSSR \cite{LFSSR}~ & ~~ 2$\times$  ~~ & ~~ 41.55/0.984 ~~ & ~~ 4$\times$  ~~ & ~~ 36.77/0.957  \\
LF-ATO \cite{ATO}~ & ~~ 2$\times$  ~~ & ~~ 41.80/0.985 ~~ & ~~ 4$\times$  ~~ & ~~ 36.95/0.959  \\
LF-InterNet ~~ & ~~ 2$\times$  ~~ & ~~ 42.36/0.985 ~~ & ~~ 4$\times$  ~~ & ~~ 37.12/0.960 \\
\hline
\end{tabular}
\vspace{-0.2cm}
\end{table}

\begin{figure}[t]
\centering
\vspace{-0.0cm}
\includegraphics[width=12.2cm]{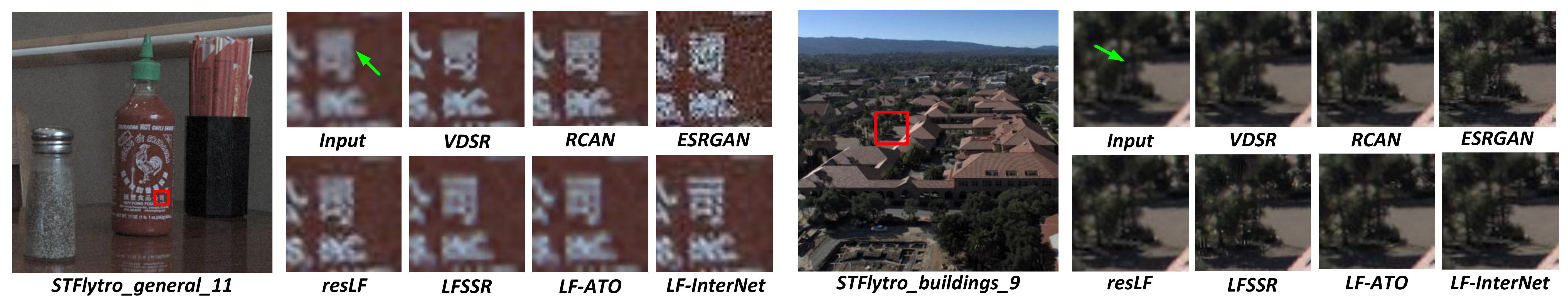}
\vspace{-0.5cm}
\caption{Visual results achieved by different methods under real-world degradation.} \label{VisualReal}
\vspace{-0.3cm}
\end{figure}

\textbf{Generalization to Unseen Scenarios.} We evaluate the generalization capability of different methods by testing them on a novel and unseen real-world dataset (i.e., the UCSD dataset \cite{UCSD}). Note that, all methods have not been trained or fine-tuned on the UCSD dataset. Results in Table \ref{tabGeneralization} show that our LF-InterNet outperforms the state-of-the-art methods \cite{RCAN,SAN,resLF,LFSSR,ATO}, which demonstrates the generalization capability of our method to unseen scenarios.

\textbf{Performance Under Real-World Degradation.} We compare the performance of different methods under real-world degradation by directly applying them to LFs in the STFlytro dataset \cite{STFlytro}. As shown in Fig.~\ref{VisualReal}, our method produces images with faithful details and less artifacts. Since the LF structure keeps unchanged under both bicubic and real-world degradation, our method can learn to incorporate spatial and angular information from training LFs using the proposed spatial-angular interaction mechanism. It is also demonstrated that our method can be easily applied to LF cameras to generate high-quality images.

\vspace{-0.2cm}
\section{Conclusion and Future Work}
\vspace{-0.2cm}
 In this paper, we proposed a deep convolutional network LF-InterNet for LF image SR. We first introduce an approach to extract and decouple spatial and angular features, and then design a feature interaction mechanism to incorporate spatial and angular information. Experimental results have demonstrated the superiority of our LF-InterNet over state-of-the-art methods. Since the spatial-angular interaction mechanism is a generic framework and can process LFs in an elegant and efficient manner, we will apply LF-InterNet to LF angular SR \cite{wu2017light,wu2019learning,jin2020learning,shi2020learning} and joint spatial-angular SR \cite{meng2019high,meng2020high} as our future work.

\section{Acknowledgement}
This work was supported by the National Natural Science Foundation of China (No. 61972435, 61602499), Natural Science Foundation of Guangdong Province, Fundamental Research Funds for the Central Universities (No. 18lgzd06).

\clearpage
\bibliographystyle{splncs}
\bibliography{LF-InterNet}

\end{document}